\documentclass[aps,prb,10pt,amsmath,amssymb,twocolumn,longbibliography,superscriptaddress]{revtex4-2}
\usepackage{hyperref}
\usepackage{graphicx,tikz}
\usepackage{mathtools}
\usepackage{array}
\usepackage{multirow}
\usepackage{orcidlink}
\usepackage{bm}

\def\be#1\ee{\begin{align}#1\end{align}}
\newcommand{\rmd}{\mathrm{d}}

\begin{document}
\title{Magnetoelastic Waves in Ferromagnetic Thin Films Mediated by Dipolar Interactions}

\author{Hiroki Yoshida\,\orcidlink{0000-0002-4505-047X}}
\affiliation{Department of Physics, Institute of Science Tokyo, 2-12-1 Ookayama, Meguro-ku, Tokyo 152-8551, Japan}

\author{Ryohei Kono}
\affiliation{Department of Physics, Institute of Science Tokyo, 2-12-1 Ookayama, Meguro-ku, Tokyo 152-8551, Japan}

\author{Manato Fujimoto\,\orcidlink{0000-0003-4882-0465}}
\affiliation{Department of Applied Physics, The University of Tokyo, 7-3-1 Hongo, Bunkyo-ku, Tokyo 113-8656, Japan}

\author{Motoki Asano\,\orcidlink{0000-0002-7059-599X}}
\affiliation{Basic Research Laboratories, NTT, Inc., 3-1 Morinosato-Wakamiya, Atsugi-shi, Kanagawa 243-0198, Japan}

\author{Daiki Hatanaka\,\orcidlink{0000-0003-0974-6207}}
\affiliation{Basic Research Laboratories, NTT, Inc., 3-1 Morinosato-Wakamiya, Atsugi-shi, Kanagawa 243-0198, Japan}

\author{Kei Yamamoto\,\orcidlink{0000-0001-9888-4796}}
\affiliation{Advanced Science Research Center, Japan Atomic Energy Agency, Tokai, Ibaraki 319-1195, Japan}
\affiliation{Center for Emergent Matter Science, RIKEN, 2-1 Hirosawa, Wako, Saitama 351-0198, Japan}

\author{Shuichi Murakami\,\orcidlink{0000-0002-2033-9402}}
\affiliation{Department of Applied Physics, The University of Tokyo, 7-3-1 Hongo, Bunkyo-ku, Tokyo 113-8656, Japan}
\affiliation{Center for Emergent Matter Science, RIKEN, 2-1 Hirosawa, Wako, Saitama 351-0198, Japan}
\affiliation{International Institute for Sustainability with Knotted Chiral Meta Matter (WPI-SKCM$^{\mathit2}$), Hiroshima University, Higashi-hiroshima, Hiroshima 739-0046, Japan}

\date{\today}

\begin{abstract}
    Magnetoelastic coupling mediated by magnetic dipolar interactions is theoretically investigated in ferromagnetic thin films under an in-plane magnetic field. We develop a theoretical description that incorporates dipolar fields derived from Maxwell's equations in the presence of elastic deformations. The resulting coupled equations of motion predict hybridization between magnetostatic and Lamb waves. Numerical calculations for a yttrium iron garnet (YIG) film reveal anti-crossings in the dispersion relations, with hybridization gaps ranging from $0.1$ to several MHz.
\end{abstract}

\maketitle

Magnetoelastic coupling has attracted considerable attention because of its potential for manipulating magnetic properties via mechanical degrees of freedom~\cite{Kittel_PR,LL_EDinCM}. At the microscopic level, magnons and phonons are coupled via spin-orbit interaction~\cite{Kittel_PR,Kittel_Abraham_RMP,Flebus_magnonPolaron}. In contrast, at longer wavelengths, magnetic dipolar interactions also play an important role for the coupling. Recent experimental advances, particularly those based on interdigital transducers (IDTs), have enabled coherent excitation and precise characterization of acoustic waves in magnetic thin films~\cite{IDT_review,Kunstle2025}, allowing direct access to magnetoelastic coupling in this regime. While continuum theories describing magnetostatic~\cite{Walker1957,damoneshbachMagnetostaticModesFerromagnet1961,damonPropagationMagnetostaticSpin} and elastic waves~\cite{Rayleigh1885,Rayleigh1889,Lamb1889,lambWavesElasticPlate1917} in this regime have been developed separately, a unified theoretical framework that captures their coupling mediated by dipolar interactions remains lacking, primarily due to the long-range nature of the dipolar interaction and its singular behavior at short distances.

In this Letter, we present a formulation of magnetoelastic coupling between magnetostatic and elastic waves mediated by dipolar interactions. Using a Green-function approach along the same lines as that developed by Kalinikos and Slavin~\cite{Kalinikos_Slavin_1986}, we consistently incorporate elastic deformations into the description, thereby enabling a quantitative evaluation of the coupling.

\begin{figure}
    \begin{center}
        \includegraphics[width=\columnwidth]{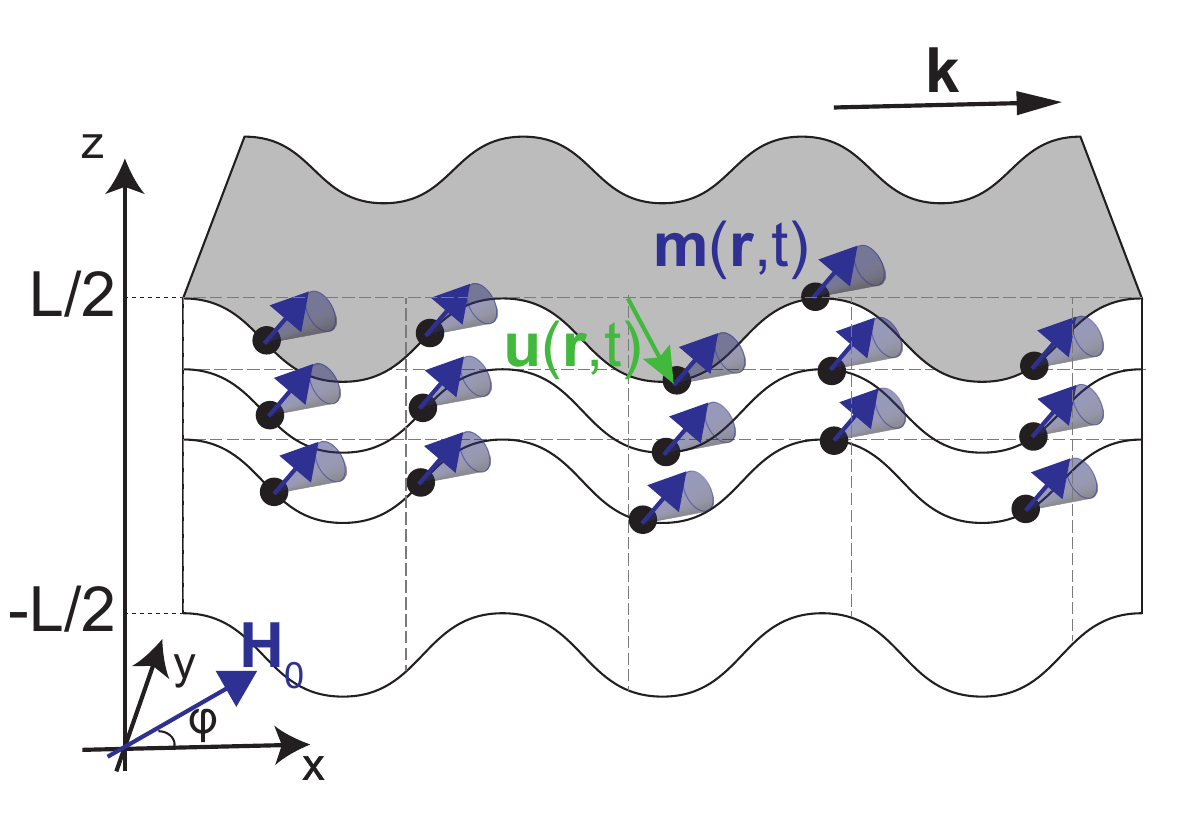}
        \caption{Schematic illustration of the system under consideration. A ferromagnetic thin film with thickness $L$ in the $z$ direction and infinite extent in the $x$ and $y$ directions is considered. Magnetic moments precess around an axis parallel to the in-plane external magnetic field $\bm{H}_0$, which is applied at an angle $\varphi$ with respect to the $x$ axis. The elastic wave changes the positions of the magnetic moments and their separations, resulting in a modification of the dipolar interaction among them.}
        \label{fig:Schematic}
    \end{center}
\end{figure}

Our system is schematically shown in Fig.~\ref{fig:Schematic}. We consider a ferromagnetic thin film that is infinitely extended in the $xy$ plane and has a finite thickness $L$ in the $z$ direction, with the origin taken at the middle of the film. It is assumed to be isotropic both magnetically and elastically. An in-plane external magnetic field $\bm{H}_0$ is applied, resulting in an equilibrium saturation magnetization $\bm{M}_0$ parallel to $\bm{H}_0$.

We decompose the magnetic field $\bm{H}$ and the magnetization $\bm{M}$ as $\bm{H}=\bm{H}_0+\bm{h}(\bm{r},t)$ and $\bm{M}=\bm{M}_0+\bm{m}(\bm{r},t)$, respectively, where $\bm{r}=(x,y,z)$ is the Euler coordinate, and study propagating magnetostatic waves of small deviations from the equilibrium magnetization. The system also supports elastic waves described by the displacement $\bm{u}(\bm{r},t)$. The deformation modulates the distance between magnetic dipoles, so that the dipolar field $\bm{h}(\bm{r},t)$ generated by $\bm{m}$ is influenced by $\bm{u}$, leading to a magnetoelastic wave. We take the $x$ axis to be parallel to the propagation direction and consider plane waves. The system is assumed to be uniform in the $y$ direction. Then, $\bm{m}$ and $\bm{u}$ can be written as
\be
    \bm{m}(\bm{r},t)&=\bm{m}(z)e^{i(kx-\omega t)},\\
    \bm{u}(\bm{r},t)&=\bm{u}(z)e^{i(kx-\omega t)},\label{eq:Assumption}
\ee
where $\bm{m}(z),\bm{u}(z)\in\mathbb{C}^3$. Hereafter, all fields are treated as complex quantities, and physical observables are given by their real parts. We study the coupled dynamics of these waves up to first order in these quantities. Here, we focus exclusively on the case of an in-plane magnetic field. The case of an arbitrary external magnetic field will be presented elsewhere.

We start from the basic equations describing the magnetic dipolar interaction, namely Maxwell's equations. When elastic deformation is present, the magnetic dipoles deviate from their equilibrium positions $\bm{r}$ to $\bm{R}\coloneqq \bm{r}+\bm{u}(\bm{r},t)$. The Maxwell's equations are then rewritten in terms of this Lagrange coordinate as
\be
    \nabla_{\bm{R}}\cdot \left(\bm{H}+\bm{M}\right)&=0\label{eq:Maxwell_R_div},\\
    \nabla_{\bm{R}}\times \bm{H}&\approx 0,\label{eq:Maxwell_R_rot}
\ee
where in the second equation the magnetostatic approximation is applied and the time derivative of the electric field is discarded. This approximation is justified in the absence of steady currents and under the quasi-static condition $\omega L / c \ll 1$, where $\omega$ and $L$ are the characteristic frequency and length scale of the magnetization dynamics and $c$ is the speed of light. In this regime, retardation effects and electromagnetic radiation can be neglected~\cite{Stancil_TheoryofMagnetostaticWaves}. By the plane-wave assumption~\eqref{eq:Assumption}, the dipolar field can also be written as $\bm{h}=\bm{h}(z)e^{i(kx-\omega t)}$. We rewrite Eqs.~\eqref{eq:Maxwell_R_div} and ~\eqref{eq:Maxwell_R_rot} in terms of these quantities in the Euler coordinate $\bm{r}$. We note that we need to pay attention to the conservation of the magnetization in a unit volume~\cite{Yamamoto_Maekawa_AnnPhys} yielding $\bm{M}(\bm{R})=\bm{M}(\bm{r})/\det(J(\bm{r}))$, where $J(\bm{r})$ is a Jacobian matrix (see Supplementary data~\cite{SM} for details). We get
\be
    \nabla_{\bm{r}}\cdot\left(\bm{h}(\bm{r})+\bm{m}(\bm{r})\right)&=\bm{M}_0\cdot\nabla_{\bm{r}}\left(\nabla_{\bm{r}}\cdot\bm{u}(\bm{r})\right),\label{eq:Maxwell_couple_div}\\
    \nabla_{\bm{r}}\times\bm{h}(\bm{r})&=0,\label{eq:Maxwell_couple_rot}
\ee
where $\nabla_{\bm{r}}$ stands for derivatives with respect to the $\bm{r}$ coordinate. Because the surface of the film is also distorted as in Fig.~\ref{fig:Schematic}, the boundary conditions used to solve these equations are also modified by the presence of $\bm{u}$ as
\be
    h^{\mathrm{in}}_x-h_x^{\mathrm{out}}&=0,\label{eq:Maxwell_bc_x}\\
    h^{\mathrm{in}}_z+m_z-h^{\mathrm{out}}_z&=M_{0,x}\frac{\partial u_z}{\partial x},\label{eq:Maxwell_bc_z}
\ee
at $z=\pm L/2$, where $h^{\mathrm{in}/\mathrm{out}}$ stand for the dipolar fields inside and outside the film, respectively. Hereafter, lower indices indicate components in the Euler coordinate. Following the Green tensor method as in Ref.~\cite{Kalinikos_Slavin_1986}, the dipolar field satisfying Eqs.~\eqref{eq:Maxwell_couple_div} and \eqref{eq:Maxwell_couple_rot} under the boundary conditions \eqref{eq:Maxwell_bc_x} and \eqref{eq:Maxwell_bc_z} is given as (see Supplementary data~\cite{SM} for details)
\be
    \bm{h}(z)=\int_{-\frac{L}{2}}^{\frac{L}{2}}\rmd z'\left(G^m(z,z')\bm{m}(z')+G^u(z,z')\bm{u}(z')\right),\label{eq:dipolar_field}
\ee
where
\be
    G^m(z,z')&=\left(\begin{array}{ccc}
        -G_P(z,z')&0&-iG_Q(z,z')\\
        0&0&0\\
        -iG_Q(z,z')&0&G_{P'}(z,z')
    \end{array}\right),\\
    G_{P}(z,z')&=\frac{k}{2}e^{-k|z-z'|},\\
    G_{Q}(z,z')&=\mathrm{sgn}(z-z')G_{P}(z,z'),\\
    G_{P'}(z,z')&=G_{P}(z,z')-\delta(z-z').
\ee
This dipolar field couples magnetostatic and elastic waves in the film. In the present setup, only elastic deformations in the $xz$ plane couple to the magnetization. In addition, the coupling vanishes when the equilibrium magnetization is parallel to the $y$ axis.

Next, we examine the effect of magnetoelastic coupling on the dynamics of the magnetostatic and elastic waves. In the absence of this coupling, the dynamics of the magnetization $\bm{m}$ up to first order is described by the linearized Landau--Lifshitz equation
\be
    \dot{\bm{m}}=-\gamma\mu_0\left(\bm{m}\times\bm{H}_0+\bm{M}_0\times\bm{h}\right),\label{eq:LandauLifshitz}
\ee
where $\gamma$ is the gyromagnetic ratio and $\mu_0$ is the vacuum magnetic permeability. This equation can be derived from a Lagrangian
\be
    L^m[\bm{m},\dot{\bm{m}},\bm{m}^*,\dot{\bm{m}}^*]=\int\rmd z\Re\left[\frac{1}{4M_0^2\gamma}\bm{M}_0\cdot(\bm{m}\times\dot{\bm{m}}^*)\right.\nonumber\\
    \hspace{2.9cm}\left.-\frac{\mu_0}{4}\left(\bm{h}^*\cdot\bm{h}+\frac{H_0}{M_0}\bm{m}^*\cdot\bm{m}\right)\right],\label{eq:Lagrangian_m}
\ee
where the asterisk denotes complex conjugate. We can verify this by constructing the Euler--Lagrange equation with respect to $\bm{m}^*$ from this Lagrangian. On the other hand, the dynamics of the displacement field $\bm{u}$ is described by the Navier--Cauchy equation
\be
    \rho\ddot{\bm{u}}=\mu\nabla^2\bm{u}+\left(\mu+\lambda\right)\nabla\left(\nabla\cdot\bm{u}\right),\label{eq:NavierCauchy}
\ee
where $\rho$ is the mass density, $\mu$ and $\lambda$ are Lam\'e constants. The Lagrangian for this equation is
\be
    L^u[\bm{u},\dot{\bm{u}},\bm{u}^*,\dot{\bm{u}}^*]=\int\rmd z\Re\left[\frac{\rho}{4}\dot{\bm{u}}^*\cdot\dot{\bm{u}}+\frac{1}{4}\partial_i\sigma_{ij}u_j^*\right],\label{eq:Lagrangian_u}
\ee
where $\sigma_{ij}=\mu(\partial_i u_j+\partial_j u_i)+\lambda\delta_{ij}\nabla\cdot\bm{u}$ is the stress tensor. The total Lagrangian of the coupled system is given by the sum of the two Lagrangians, Eqs.~\eqref{eq:Lagrangian_m} and \eqref{eq:Lagrangian_u}. In this case, $\bm{h}$ is no longer only a functional of the magnetization, but also a functional of the displacement field $\bm{u}$. As a result,  a magnetoelastic coupling term of the form
\begin{equation}
    L^{\mathrm{me}}
    =
    \frac{\mu_0 k}{2}
    \iint\mathrm{d}z'\mathrm{d}z''\,
    \Im\left[
    M_{0,x}\bm{m}^*(z')\cdot G^m(z',z'')\bm{u}(z'')
    \right]
\end{equation}
arises from the $\bm{h}^*\cdot\bm{h}$ term in the Lagrangian. This term is analogous to the standard local magnetoelastic coupling density~\cite{Kittel1949,KittelAbrahams1953,Neel1954, CallenCallen1963}, while explicitly retaining the long-range character of the dipolar interaction. In addition, the elasticity tensor receives a magnetostatic contribution, a new volume force term $-\bm{f}$ appears in the Navier--Cauchy equation~\eqref{eq:NavierCauchy}, where the force density $\bm{f}$ is given by (see Supplementary data~\cite{SM} for details of the derivation)
\be
    f_i(z)\coloneqq \mu_0\frac{\delta h^*_j}{\delta u^*_i}h_j=-ik\mu_0 M_{0,x}h_i(z).
\ee
Physically, this force can be understood as a Maxwell stress acting on the magnetic dipoles, caused by the change in the dipolar field induced by the displacement field. The coupled dynamics of the magnetoelastic wave is then given by
\be
    \left\{
    \begin{array}{l}
        \dot{\bm{m}}=-\gamma\mu_0\left(\bm{m}\times\bm{H}_0+\bm{M}_0\times\bm{h}\right)\\
        \rho\ddot{\bm{u}}=\mu\nabla^2\bm{u}+\left(\mu+\lambda\right)\nabla\left(\nabla\cdot\bm{u}\right)-\bm{f}.
    \end{array}\label{eq:CoupledEOM}
    \right.
\ee

\begin{figure*}[t]
    \begin{center}
        \includegraphics[width=2 \columnwidth]{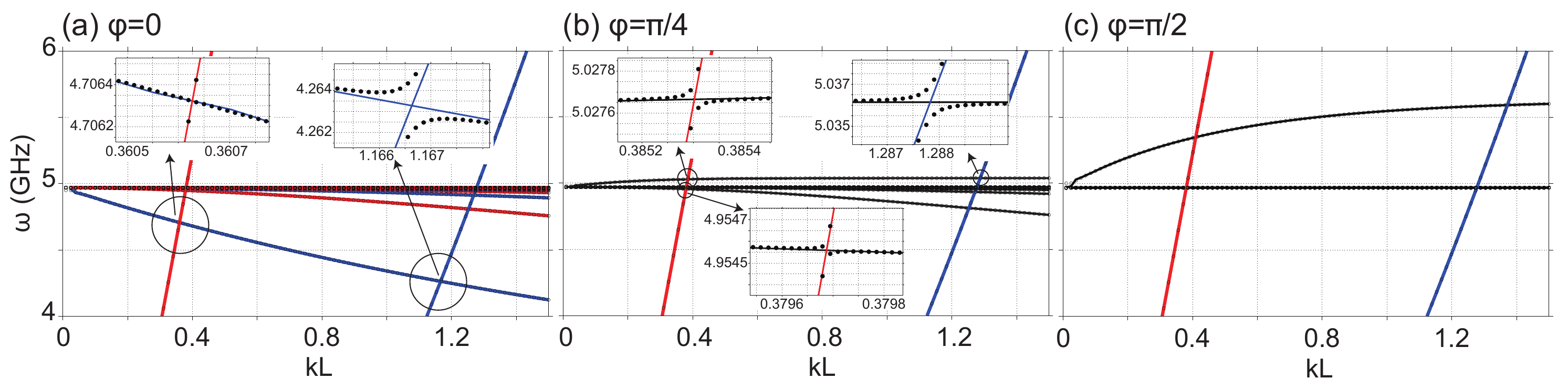}
        \caption{Dispersion curves of the coupled waves. Red and blue curves represent modes with $C_{2x}$ eigenvalues $+1$ and $-1$, respectively. Black curves represent magnetostatic modes that are not eigenstates of the $C_{2x}$ operator. Markers show the dispersion of the coupled waves obtained from numerical calculations. The external magnetic field is applied (a) parallel to the propagation direction, $\bm{H}_0 \parallel \bm{k}$ ($\varphi=0$), (b) at $\varphi=\pi/4$, and (c) perpendicular to the propagation direction, $\bm{H}_0 \perp \bm{k}$ ($\varphi=\pi/2$). Insets show magnified views of representative band crossings. In (a), $\varphi=0$, anticrossings appear between BVW and Lamb modes only when they have the same $C_{2x}$ eigenvalue (same color). When the $C_{2x}$ symmetry of the system is broken (b), both the magnetostatic surface mode and the BVW couple to Lamb modes. The bottom-center inset of panel (b) shows the hybridization between the lowest-frequency BVW mode and the first Lamb wave mode.}
        \label{fig:Bands}
    \end{center}
\end{figure*}

In order to examine the dynamical properties of the coupled waves described by the above equations of motion, we numerically calculate the dispersion relation in the following. Under the assumption~\eqref{eq:Assumption}, the time derivatives on the left-hand side of the equations become $-i\omega$. Since we consider a film in vacuum, we impose the stress-free condition at the surfaces $z=\pm L/2$ as
\be
    \sigma_{xz}|_{z=\pm\frac{L}{2}}=\sigma_{yz}|_{z=\pm\frac{L}{2}}=\sigma_{zz}|_{z=\pm\frac{L}{2}}=0.\label{eq:BC_EW}
\ee
Without the coupling, it is known that a type of elastic wave called a Lamb wave appears under these conditions. Since the system in equilibrium is symmetric with respect to the $z=0$ plane, the Lamb wave modes are either even or odd in $z$. For later convenience, we introduce the $C_{2x}$ symmetry operation. The system is invariant under $C_{2x}$, and the deformation vector $\bm{u}(z)$ is an eigenfunction of the corresponding operator: $C_{2x}\bm{u}(z)=\pm\bm{u}(z)$. Numerically, we need to solve the equations under the boundary conditions~\eqref{eq:BC_EW}. For example, the finite element method naturally incorporates these conditions.

On the other hand, the magnetostatic waves without the coupling to elastic waves in this geometry were studied in detail by Damon and Eshbach~\cite{damoneshbachMagnetostaticModesFerromagnet1961}. They analytically showed that when the external magnetic field is parallel to the propagation direction of the wave ($\bm{H}_0\parallel\bm{k}$), the system has infinitely many bulk modes called backward volume waves (BVWs). When the magnetic field is in-plane but perpendicular to the propagation direction ($\bm{H}_0\perp\bm{k}$), a surface mode appears that is localized at one of the surfaces.

For the calculation of the dispersion, it is convenient to introduce another coordinate system, where the original one $(x,y,z)$ is rotated around the $z$ axis so that one of the axes is parallel to the external magnetic field. Then, the component of $\bm{m}$ is zero along this direction since $\bm{M}_0\cdot\bm{m}=0$ from the constraint $|\bm{M}|=const.$

Now we consider the coupled equations~\eqref{eq:CoupledEOM}. Using the finite element method, we discretize the equations in real space while satisfying the boundary conditions, which leads to a generalized eigenvalue problem for the eigenvalue $\omega$. By solving this problem, we obtain the dispersion curves shown in Fig.~\ref{fig:Bands}. The parameters are taken to be those of Yttrium-Iron-Garnet (YIG): $\rho=5.17\ \mathrm{g}/\mathrm{cm}^3$, $\mu=7.81\times10^{-8}\ \mathrm{kg}\cdot\mathrm{GHz}^2/\mathrm{m}$, $\lambda=10.94\times10^{-8}\ \mathrm{kg}\cdot\mathrm{GHz}^2/\mathrm{m}$~\cite{clarkElasticConstantsSingleCrystal1961}, $\gamma=29.8\ \mathrm{GHz}/\mathrm{T}$, and $\mu_0M_0=0.178$ T~\cite{AulockHandbookofferritematerials}. The vacuum permeability is $\mu_0=4\pi\times10^{11}\ \mathrm{T}^2\mathrm{m}/(\mathrm{kg}\cdot\mathrm{GHz}^2)$. We set the film thickness to $L=500$ nm and the external magnetic field to $\mu_0H_0=0.1$ T.

We define the angle between the $x$ axis and the external magnetic field as $\varphi$ and write $\bm{H}_0=H_0(\cos\varphi,\sin\varphi,0)$. In Fig.~\ref{fig:Bands}(a), we plot the case $\varphi=0$. In this case, the system has a $C_{2x}$ rotation symmetry and thus, the modes of magnetostatic BVWs and Lamb waves can be labeled by their eigenvalues. States with $C_{2x}$ eigenvalue $-1$ are plotted as blue curves, while those with $+1$ as red curves. Curves with negative slopes are magnetostatic BVWs and positive ones are Lamb waves, respectively. The markers represent the dispersion of the coupled waves. The insets show magnified views of the band crossings between BVWs and Lamb waves. We observe that the magnetostatic and elastic waves cause anti-crossing and produce hybridization gaps of several MHz only when the two modes have the same $C_{2x}$ eigenvalues. Figure~\ref{fig:Bands}(b) shows the case $\varphi=\pi/4$. In this case, two types of magnetostatic modes coexist: BVWs and surface waves. The insets show magnified views of the band crossings between the magnetostatic and Lamb waves. Hybridization gaps at anti-crossings appear between all the modes as the system does not have any symmetries in this case. Finally, in Fig.~\ref{fig:Bands}(c), we plot the case $\varphi=\pi/2$. Here, a magnetostatic surface mode appears but does not couple to the Lamb waves. This is because $G^u=0,\bm{f}=\bm{0}$, so that Eq.~\eqref{eq:CoupledEOM} reduces to two independent equations. The decoupling arises from the mirror symmetry with respect to the plane normal to the $y$ axis, under which the magnetostatic and Lamb waves are odd and even, respectively. This symmetry argument does not constrain the coupling to deformations along the $y$ direction. Nevertheless, Maxwell's equations~\eqref{eq:Maxwell_couple_div} and \eqref{eq:Maxwell_couple_rot} do not involve $u_y$ within the magnetostatic approximation and to linear order in $\bm{m}$ and $\bm{u}$. Therefore, in the present theory, magnetostatic and elastic waves are completely decoupled when $\varphi=\pi/2$.

From the magnitude of the hybridization gaps, the effective coupling strength $g_{\mathrm{eff}}$ is estimated to be on the order of a few MHz. On the other hand, assuming a typical Gilbert damping parameter $\alpha \approx 10^{-4}$~\cite{YIGalpha_PRAppl}, a quality factor $Q \approx 10^3$, and operating frequencies $\omega$ in the GHz range, the corresponding magnon and phonon linewidths are estimated as $\gamma_m \sim \alpha \omega$ and $\kappa \sim \omega/Q$, respectively. Under these conditions, $g_{\mathrm{eff}}^2/(\gamma_m\kappa)\sim1$, suggesting that the system is close to the boundary of the strong-coupling regime~\cite{rodriguezClassicalQuantumDistinctions2016}. Therefore, the hybridization gaps are expected to be experimentally resolvable.

To generate the same gap via magnetoelastic coupling for bulk transverse waves, the magnetoelastic coupling constant $b_2$ would need to be on the order of $10^4$ J/m$^3$~\cite{Hioki2022}, which is one order of magnitude smaller than the experimental value for YIG, $b_2 \sim 10^5$ J/m$^3$~\cite{Krysztofik2021}. This result indicates that the dipolar contribution to the magnetoelastic coupling is subdominant compared to other mechanisms in realistic systems.

In conclusion, we have developed a theoretical description of magnetoelastic coupling mediated by magnetic dipolar interactions in ferromagnetic thin films. The dipolar field was derived from Maxwell's equations in the presence of elastic deformations, and the coupled equations of motion were obtained within the Lagrangian formalism. Numerical calculations were performed for a YIG thin film to obtain the dispersion relations of the coupled modes. We found anti-crossings due to mode hybridization, with gaps of several MHz in general. When the external in-plane magnetic field is parallel to the wave propagation, the system has $C_2$ symmetry, so only magnetostatic and elastic waves with the same $C_2$ symmetry eigenvalues can hybridize. When they are perpendicular, couplings do not occur. The formulation developed here provides a starting point for theoretical studies of dipolar-interaction-mediated magnetoelastic coupling in ferromagnetic thin films.

\textit{acknowledgements}--H. Y. acknowledges support from Japan Society for the Promotion of Science (JSPS) KAKENHI Grant Number JP24KJ1109 and MEXT Initiative to Establish Next-generation Novel Integrated Circuits Centers (X-NICS) Grant Number JPJ011438. 
M. F. acknowledges support from Japan Society for the Promotion of Science (JSPS) KAKENHI Grant Numbers JP23KJ0339 and JP24K16987, and JST PRESTO Grant Number JPMJPR25H8. 
M. A. acknowledges support from Japan Society for the Promotion of Science (JSPS) KAKENHI Grant Number JP23H05463. 
D. H. acknowledges support from Japan Society for the Promotion of Science (JSPS) KAKENHI Grant Numbers JP21H05020, JP24H02235, and JP23H05463. 
K. Y. acknowledges support from Japan Society for the Promotion of Science (JSPS) KAKENHI Grant Numbers JP24K00576 and JP25H00837 and from JSPS Bilateral Program Number JPJSBP120245708. 
S. M. acknowledges support from Japan Society for the Promotion of Science (JSPS) KAKENHI Grant Numbers JP22K18687, JP22H00108, JP24H02230, and JP24H02231.

\bibliography{Ref_MSEW.bib}

\clearpage
\onecolumngrid
\setcounter{equation}{0}
\renewcommand{\theequation}{S\arabic{equation}}

\section*{Magnetoelastic Waves in Ferromagnetic Thin Films Mediated by Dipolar Interactions: Supplementary data}


\subsection{Jacobian matrix}
\label{sec:Jacobian}

Here we calculate the Jacobian matrix and its determinant for the transformation $\bm{r}\mapsto\bm{R}=\bm{r}+\bm{u}(\bm{r},t)$ up to first order in $\bm{u}$.

The definition of the Jacobian matrix $J$ is
\be
    J_{ij}(\bm{r})&\coloneqq \delta_{ij}+\frac{\partial u_i}{\partial r_j}.
\ee
Then, its determinant is given as
\be
    \det J&=\varepsilon_{i_1i_2i_3}J_{1i_1}J_{2i_2}J_{3i_3}\nonumber\\
    &=\varepsilon_{i_1i_2i_3}\left(\delta_{1i_1}+\frac{\partial u_{1}}{\partial r_{i_1}}\right)\left(\delta_{2i_2}+\frac{\partial u_{2}}{\partial r_{i_2}}\right)\left(\delta_{3i_3}+\frac{\partial u_{3}}{\partial r_{i_3}}\right)\nonumber\\
    &=1+\frac{\partial u_{1}}{\partial r_{1}}+\frac{\partial u_{2}}{\partial r_{2}}+\frac{\partial u_{3}}{\partial r_{3}}+\mathcal{O}\left((\partial u)^2\right)\nonumber\\
    &\approx 1+\nabla\cdot\bm{u}.
\ee
For a nonsingular $3\times3$ matrix \(A\), its inverse is given by
\be
    A^{-1}&=\frac{1}{\det A}C^T,
\ee
where \(C\) is a cofactor matrix defined by
\be
    C_{ij}&\coloneqq \frac{1}{2}\varepsilon_{ii_1i_2}\varepsilon_{jj_1j_2}A_{i_1j_1}A_{i_2j_2}.
\ee
Applying this to the Jacobian matrix, we obtain
\be
    \left[\left(J^{-1}\right)^T\right]_{ij}&=\left(J^{-1}\right)_{ji}\nonumber\\
    &=\frac{1}{\det J}C_{ij}\nonumber\\
    &=\frac{1}{2\det J}\varepsilon_{ii_1i_2}\varepsilon_{jj_1j_2}J_{i_1j_1}J_{i_2j_2}\nonumber\\
    &=\frac{1}{2\det J}\varepsilon_{ii_1i_2}\varepsilon_{jj_1j_2}\left(\delta_{i_1j_1}+\frac{\partial u_{i_1}}{\partial r_{j_1}}\right)\left(\delta_{i_2j_2}+\frac{\partial u_{i_2}}{\partial r_{j_2}}\right)\nonumber\\
    &\approx \frac{1}{2}\left(1-\nabla\cdot\bm{u}\right)\left(\varepsilon_{ii_1i_2}\varepsilon_{ji_1i_2}+\varepsilon_{ij_1i_2}\varepsilon_{jj_1j_2}\frac{u_{i_2}}{r_{j_2}}+\varepsilon_{ii_1j_2}\varepsilon_{jj_1j_2}\frac{u_{i_1}}{r_{j_1}}\right)\nonumber\\
    &=\left(1-\frac{\partial u_{i_1}}{\partial r_{i_1}}\right)\delta_{ij}+\frac{1}{2}\left(1-\frac{\partial u_{i_1}}{\partial r_{i_1}}\right)\left(2\delta_{ij}\frac{\partial u_{i_1}}{\partial r_{i_1}}-2\frac{\partial u_j}{\partial r_i}\right)\nonumber\\
    &\approx\delta_{ij}-\frac{\partial u_j}{\partial r_i}.
\ee

\subsection{Derivation of the dipolar field}
\label{sec:Derivation_h}

Here, we derive the dipolar field that gives rise to magnetoelastic coupling.

We start from Maxwell's equations in the Lagrangian coordinate $\bm{R}$:
\be
    \nabla_{\bm{R}}\cdot(\bm{H}_{\mathrm{in}}(\bm{R},t)+\bm{M}(\bm{R},t))&=0,\\
    \nabla_{\bm{R}}\times\bm{H}_{\mathrm{in}}(\bm{R},t)&\approx0,
\ee
where \(\nabla_{\bm{R}}\) denotes the derivative with respect to \(\bm{R}\). We then rewrite these equations in terms of the Eulerian coordinate $\bm{r}$. In doing so, we must take into account the conservation of magnetization~\cite{Yamamoto_Maekawa_AnnPhys}:
\be
    \int_{V_R}\rmd^3R\,\bm{M}(\bm{R},t)&=\int_{V_r}\rmd^3r\, \bm{M}(\bm{r},t)\nonumber\\
    \Rightarrow \int_{V_r}\rmd^3r\,(\det J)\bm{M}(\bm{R}(\bm{r}),t)&=\int_{V_r}\rmd^3r\, \bm{M}(\bm{r},t)\qquad \forall V_r.\nonumber\\
    \therefore \bm{M}(\bm{R})&=\frac{1}{\det J(\bm{r})}\bm{M}(\bm{r}).
\ee
Then, using the Jacobian matrix of the coordinate transformation, the Maxwell's equations become
\be
    \nabla_{\bm{R}}\cdot(\bm{H}_{\mathrm{in}}(\bm{R},t)+\bm{M}(\bm{R},t))&=0\nonumber\\
    \Rightarrow\left[\left(J^{-1}\right)^T\right]_{ij}\frac{\partial}{\partial r_j}\left(h_i(\bm{r},t)+\frac{1}{\det J}M_i(\bm{r})\right)&=0,\label{eq:Maxwell_r1}
\ee
and
\be
    \nabla_{\bm{R}}\times\bm{H}_{\mathrm{in}}(\bm{R},t)&\approx0\nonumber\\
    \Rightarrow\varepsilon_{ijk}\left[\left(J^{-1}\right)^T\right]_{il}\frac{\partial}{\partial r_l}h_j(\bm{r})&\approx0\qquad\forall k=x,y,z\label{eq:Maxwell_r2}.
\ee
From the results of Sec.~\ref{sec:Jacobian}, up to first order in $\bm{m}$ and $\bm{u}$, we obtain
\be
    \nabla_r\cdot\left(\bm{h}^{\mathrm{in}}(\bm{r})+\bm{m}(\bm{r})\right)&=\bm{M}_0\cdot\nabla_r\left(\nabla_r\cdot\bm{u}(\bm{r})\right),\\
    \nabla_r\times\bm{h}^{\mathrm{in}}(\bm{r})&=0,
\ee
for points inside the thin film. The equations outside of the film are the same as those in the previous section:
\be
    \nabla\cdot\bm{h}^{\mathrm{out}}(\bm{r},t)&=0,\nonumber\\
    \nabla\times\bm{h}^{\mathrm{out}}(\bm{r},t)&\approx0.\nonumber
\ee
Since $\bm{h}^{\mathrm{in/out}}$ are rotation-free even in the presence of magnetoelastic coupling under the magnetostatic approximation, we introduce scalar potential as
\be
    \bm{h}^{\mathrm{in/out}}=-\nabla\phi^{\mathrm{in/out}}.\nonumber
\ee
Outside the film, the solution is given by
\be
    \phi^{\mathrm{out}}=\left\{\begin{array}{l}
        Ae^{+kz}\quad (z<-\frac{L}{2})\\
        Be^{-kz}\quad (z>+\frac{L}{2}),
    \end{array}\right.\label{eq:phi_out}
\ee
because $\phi^{\mathrm{out}}\to0$ as $|z|\to\infty$.

The boundary conditions are expressed in terms of the continuity of the tangential component of the magnetic field and the normal component of the magnetic flux density:
\be
    \bm{H}^{\mathrm{in}}\cdot\bm{t}&=\bm{H}^{\mathrm{out}}\cdot\bm{t},\label{eq:BC_H_tangental}\\
    \left(\bm{H}^{\mathrm{in}}+\bm{M}\right)\cdot\bm{n}&=\bm{H}^{\mathrm{out}}\cdot\bm{n},\label{eq:BC_B_normal}
\ee
where $\bm{t}$ and $\bm{n}$ are the tangential and normal vectors to the surface, respectively. Since the system is uniform in the $y$ direction, we restrict our analysis to the $xz$ plane. In this case, these vectors are given by
\be
    \bm{t}&=\left(1+\frac{\partial u_x}{\partial x},0,\frac{\partial u_z}{\partial x}\right),\\
    \bm{n}&=\left(-\frac{\partial u_z}{\partial x},0,1+\frac{\partial u_x}{\partial x}\right).
\ee
Using the solution~\eqref{eq:phi_out}, we obtain the following boundary condition for $\phi^{\mathrm{in}}$:
\be
    \left(\pm k+\partial_z\right)\phi^{\mathrm{in}}\lvert_{z=\pm\frac{L}{2}}&=\left(m_z-ikM_{0,x}u_z\right)\lvert_{z=\pm\frac{L}{2}},\label{eq:BC_coupled_phi^in}
\ee
where the signs correspond to the boundaries at $z=\pm L/2$. Accordingly, we need to solve the differential equation
\be
    \nabla^2\phi^{\mathrm{in}}&=\nabla\cdot\bm{m}-\bm{M}_0\cdot\nabla\left(\nabla\cdot\bm{u}(\bm{r})\right),
\ee
subject to the boundary condition~\eqref{eq:BC_coupled_phi^in}. 

To this end, we introduce a Green's function $g(z,z')$ satisfying
\be
    \nabla^2g(z,z')&=\delta(z-z'),\qquad \left(z,z'\in\left(-\frac{L}{2},\frac{L}{2}\right)\right)\label{eq:Coupled_g_PDE},
\ee
with the boundary condition
\be
    \left(\pm k+\partial_z\right)g(z,z')\lvert_{z=\pm\frac{L}{2}}&=0\label{eq:Coupled_g_BC}.
\ee
Using this $g(z,z')$, $\phi^{\mathrm{in}}$ can be written as
\be
    \phi^{\mathrm{in}}(z)&=\int\rmd z'\left[\nabla g(z,z')\cdot\bm{m}(z')-\nabla\left(\bm{M}_0\cdot\nabla g(z,z')\right)\cdot\bm{u}(z')\right].
\ee
The solution of Eqs.~\eqref{eq:Coupled_g_PDE} and~\eqref{eq:Coupled_g_BC} is obtained, for example, by Fourier transformation as
\be
    g(z,z')&=-\frac{1}{2k}e^{-k|z-z'|}.
\ee
Taking the gradient, we obtain the dipolar field (as given in the main text):
\be
    \bm{h}^{\mathrm{in}}(z)=\int_{-\frac{L}{2}}^{\frac{L}{2}}\rmd z'\left[G^m(z,z')\bm{m}(z')+G^u(z,z')\bm{u}(z')\right],
\ee
where the first term is the same as that in Ref.~\cite{Kalinikos_Slavin_1986}, and
\be
    G^m(z,z')&=\left(\begin{array}{ccc}
        -G_P(z,z')&0&-iG_Q(z,z')\\
        0&0&0\\
        -iG_Q(z,z')&0&G_{P'}(z,z')
    \end{array}\right),\\
    G^u(z,z')&=-ikM_{0,x}G^m(z,z'),\\
    G_{P}(z,z')&=\frac{k}{2}e^{-k|z-z'|},\\
    G_{Q}(z,z')&=\mathrm{sgn}(z-z')G_{P}(z,z'),\\
    G_{P'}(z,z')&=G_{P}(z,z')-\delta(z-z').
\ee

\subsection{Derivation of the force}
\label{sec:Derivation_f}

Here, we derive the force induced by magnetoelastic coupling.

The Euler--Lagrange equation with respect to $u^*_i$ yields
\be
    \int \rmd z\Re\left[\rho\ddot{u}_i-\left(-\mu_0\frac{\delta h_a^*}{\delta u_i^*}h_a+\partial_i\sigma_{ij}\right)\right]=0\label{eq:Formal_NC},
\ee
where the second term represents the volume force arising from the dipolar field.

A straightforward calculation shows that
\be
    \int^{\infty}_{-\infty}\rmd z''\left(G^m(z'',z)\right)^*G^m(z'',z')=\int^{\infty}_{-\infty}\rmd z''G^m(z,z'')G^m(z'',z')&=-G^m(z,z'),\label{eq:Formula_Gmprod}\\
    \int^{\infty}_{-\infty}\rmd z''\left(G^m(z'',z)\right)^*G^u(z'',z')=\int^{\infty}_{-\infty}\rmd z''G^m(z,z'')G^u(z'',z')&=-G^u(z,z').\label{eq:Formula_Gmuprod}
\ee
Then,
\be
    \int\rmd z\frac{\delta h^*_a(z)}{\delta u^*_i}h_a(z)&=\int\rmd z\rmd z'\rmd z''(+ikM_{0,x})\left(G^m(z,z')\right)^*_{ia}\left[G^m(z,z'')\bm{m}(z'')+G^u(z,z'')\bm{u}(z'')\right]_a\nonumber\\
    &=-ikM_{0,x}\int\rmd z'\int\rmd z''\left[G^m(z',z'')\bm{m}(z'')+G^u(z',z'')\bm{u}(z'')\right]_i\nonumber\\
    &=-ikM_{0,x}\int\rmd z'h^{\mathrm{in}}_i(z').
\ee
Thus, we obtain
\be
    f_i(z)\coloneqq-ik\mu_0M_{0,x}h_i(z).
\ee

\end{document}